\documentclass[9pt,conference]{IEEEtran}
\IEEEoverridecommandlockouts

\usepackage[utf8]{inputenc}
\usepackage{graphicx}
\usepackage{bm}
\usepackage{cite}
\usepackage{amsmath,amssymb,amsfonts,mathrsfs}
\usepackage{bbm}
\usepackage{amsxtra}
\usepackage{threeparttable}
\usepackage{cite}
\usepackage{mathrsfs}
\usepackage{algorithm,algorithmic}
\usepackage{etoolbox,siunitx}
\robustify\bfseries
\usepackage{multirow, booktabs}
\usepackage{balance}
\usepackage{textcomp}
\usepackage{xcolor}
\usepackage{url}
\usepackage{pifont}
\usepackage{hyperref}
\PassOptionsToPackage{hyphens}{url}

\DeclareMathOperator*{\argmin}{argmin}

\begin{document}
\hyphenpenalty=0
\linepenalty=999    

\title{Retrieval-Augmented Neural Field\\for HRTF Upsampling and Personalization
\thanks{This work was performed while C.~Ick was an intern at MERL.}
}

\author{
    \IEEEauthorblockN{\textit{
        Yoshiki Masuyama\IEEEauthorrefmark{1}, %
        Gordon Wichern\IEEEauthorrefmark{1}, %
        François G.\ Germain\IEEEauthorrefmark{1}, %
        Christopher Ick\IEEEauthorrefmark{1}\IEEEauthorrefmark{2}, %
        and Jonathan Le Roux\IEEEauthorrefmark{1}}
        \vspace{.7\baselineskip}}
        \IEEEauthorblockA{
        \IEEEauthorrefmark{1}Mitsubishi Electric Research Laboratories (MERL), Cambridge, MA, USA}
        \IEEEauthorblockA{
        \IEEEauthorrefmark{2}Music and Audio Research Laboratory, New York University, Brooklyn, NY, USA}
}


\maketitle
\begin{abstract}
Head-related transfer functions (HRTFs) with dense spatial grids are desired for immersive binaural audio generation, but their recording is time-consuming.
Although HRTF spatial upsampling has shown remarkable progress with neural fields, spatial upsampling only from a few measured directions, e.g., 3 or 5 measurements, is still challenging.
To tackle this problem, we propose a retrieval-augmented neural field (RANF).
RANF retrieves a subject whose HRTFs are close to those of the target subject from a dataset.
The HRTF of the retrieved subject at the desired direction is fed into the neural field in addition to the sound source direction itself.
Furthermore, we present a neural network that can efficiently handle multiple retrieved subjects, inspired by a multi-channel processing technique called transform-average-concatenate.
Our experiments confirm the benefits of RANF on the SONICOM dataset, and it is a key component in the winning solution of Task 2 of the listener acoustic personalization challenge 2024.
\end{abstract}


\section{Introduction}
\label{sec:intro}

Spatial audio rendering has been extensively studied for various applications, including telepresence systems~\cite{Keyrouz2007} and mixed reality technologies~\cite{Zotkin2004,Xie2013}.
To generate high-quality immersive audio over headphones, the spatial filtering effects as sound travels from its source to each ear should be mimicked by head-related transfer functions (HRTFs).
These filtering effects are caused by the reflection and scattering of sound on the upper torso, head, and pinnae.
Hence, HRTFs are different for each subject, and the use of a generic HRTF may result in suboptimal perceptual localization~\cite{Wenzel1993jasa,Oberem2020}.
Although individual HRTFs should ideally be measured with dense spatial grids, this requires an enormous amount of time and effort~\cite{Watanabe2014}.
To mitigate this problem, various HRTF spatial upsampling and personalization methods have been developed~\cite{Pulkki1997,Franck2017,Duraiswami2004,Ahrens2012,Arend2021}.

Neural networks have recently been applied to spatial upsampling, motivated by their adaptive and powerful modeling capabilities~\cite{Mo2017,Zhang2020,Ito2022,hogg2023,Gebru2021,Lee2023,Zhang2023,niirf,Etienne2024}.
In particular, neural field (NF)-based methods can exploit a dataset with arbitrary spatial grids and have shown promising performance by representing HRTFs as a function of the sound source direction~\cite{Gebru2021,Lee2023,Zhang2023,niirf}.
NF was first used to represent the HRTFs of a single subject, then extended to represent those of multiple subjects by incorporating a small number of subject-specific parameters~\cite{Zhang2023,niirf}.
As a result, a pre-trained generic NF can be easily adapted to a new subject by optimizing the subject-specific parameters while freezing most of the model parameters.
While this adaptation works well when a large number of measurements are available, spatial upsampling from fewer than ten measurements remains highly challenging.

As an orthogonal direction to spatial upsampling, HRTF selection has also been explored to synthesize plausible binaural audio.
From a given dataset with a dense spatial grid, this method aims to select a subject whose HRTFs are close to HRTFs of the target subject.
In \cite{Iwaya2006,Katz2012}, a subject manually selects the best-fitting subject from a dataset through listening tests.
To reduce the effort of the target subject, objective selection methods without human evaluation have been developed by using the similarities of HRTFs at a limited number of directions~\cite{hogg2023} or anthropometric measurements~\cite{Zotkin2004,schonstein2010,Spagnol2020}.
These methods can achieve moderate performance with a limited number of HRTF recordings or even only with anthropometric measurements.

HRTF selection motivates us to enhance NF for HRTF spatial upsampling by using selected HRTFs.
Retrieval-augmented generation (RAG)~\cite{lewis2020} has shown remarkable success in various modalities including natural language processing~\cite{lewis2020,guu2020}, computer vision~\cite{blattmann2022,horita2024}, and audio processing~\cite{Yuan2024,Mingqiu2024}.
RAG retrieves relevant information from external resources and feeds it into a neural network as a context for data generation.
This can improve the generalization capability to out-of-distribution data and the performance without increasing the model parameters.
We can interpret the selected HRTFs as the context for generating HRTFs of the target subject.

\begin{figure}[t]
    \centering
    \includegraphics[width=0.99\linewidth]{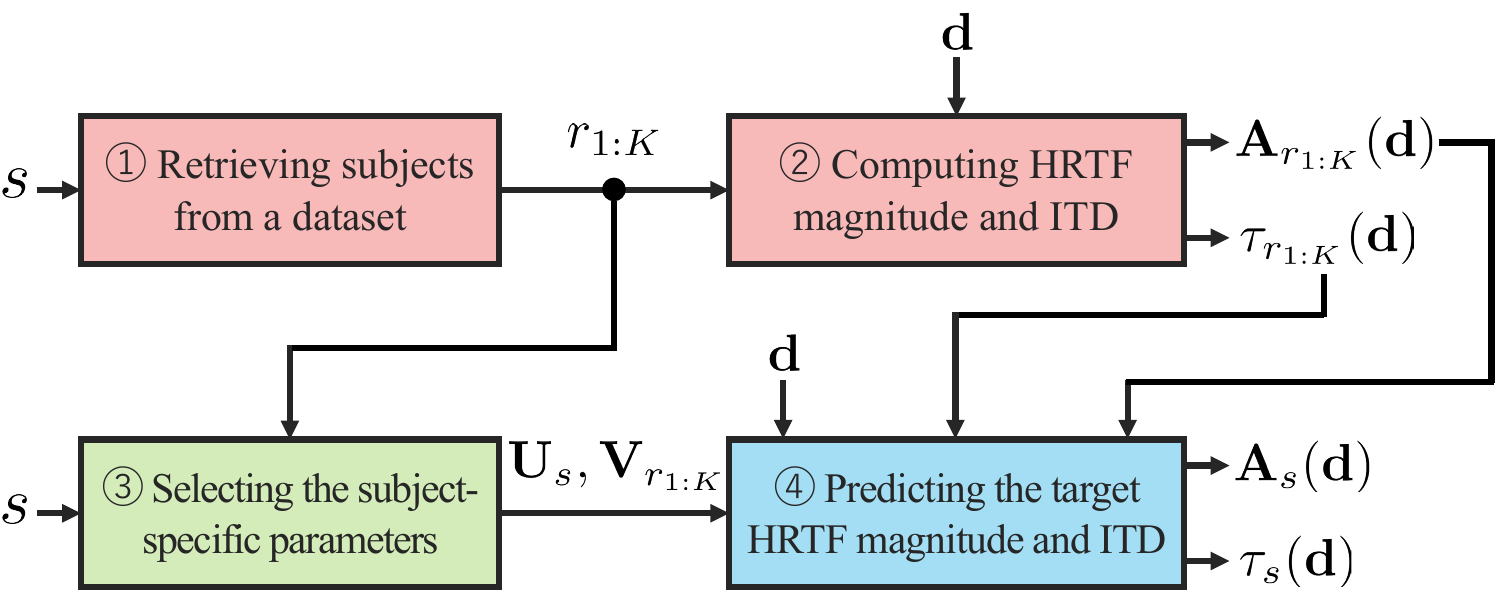}
    \caption{Overview of RANF.
    It aims to predict the HRTF magnitude and ITD of a target subject $s$ at a desired direction $\mathbf{d}$ as in \textcircled{\scriptsize 4}.
    We retrieve $K$ subjects in \textcircled{\scriptsize 1} and compute their HRTF magnitude and ITD at the desired direction $\mathbf{d}$ in \textcircled{\scriptsize 2}.
    The retrieved HRTF magnitude and ITD are fed into NF with the sound source direction and the subject-specific parameters selected in \textcircled{\scriptsize 3}.
    }
    \label{fig:overview}
    \vskip -0.1in
\end{figure}

In this paper, we propose a retrieval-augmented NF (RANF) to address the challenges in HRTF spatial upsampling from sparse measurements.
The overview of RANF is depicted in Fig.~\ref{fig:overview}.
Inspired by HRTF selection, RANF retrieves multiple subjects whose HRTFs are close to those of the target subject at the measured directions.
Then, RANF uses HRTFs of the retrieved subjects in the desired direction as an auxiliary input to NF.
Furthermore, we propose a network architecture that can efficiently exploit multiple retrievals to improve the HRTF prediction performance.
In our experiments on the SONICOM dataset~\cite{engel2023}, RANF outperformed the NF without retrieval and the HRTF selection methods.

\section{Preliminaries}

\subsection{Problem setting}
\label{sec:setting}

Let $\mathbf{H}_{s}(\mathbf{d}) \in \mathbb{C}^{F\times 2}$ denote the HRTF for a subject $s$ and direction $\mathbf{d} = (\theta, \phi)$, where $F$ is the number of frequency bins, $\theta \in [0, 2\pi)$ is the azimuth, and $\phi \in [-\pi/2, \pi/2]$ is the elevation.
The azimuth increases counter-clockwise with $\theta = 0$ directly in front of the subject, while the elevation increases upward with $\phi = 0$ corresponding to the equatorial plane.
Let $\mathcal{S}$ be the training set of subjects whose HRTFs are measured at all directions $\mathbf{d} \in \mathcal{D}$.
Meanwhile, HRTFs of a new subject $\bar{s} \notin \mathcal{S}$ are measured at a limited number of directions $\mathbf{d} \in \mathcal{D}'$, where $\mathcal{D}' \subset \mathcal{D}$.
We aim to predict $\mathbf{H}_{\bar{s}}(\mathbf{d})$ for a new subject $\bar{s}$ and all directions $\mathbf{d} \in \mathcal{D}$ from only the measurements for $\mathbf{d} \in \mathcal{D}'$.

Existing studies have shown that azimuth localization mainly relies on the interaural time difference (ITD) and the interaural level difference (ILD)~\cite{Carlini2024}, while elevation localization depends on spectral coloration~\cite{Geronazzo2018}.
Hence, HRTF phase is not essential to localization as long as ITD is preserved.
This motivates us to approximate the complex HRTF $\mathbf{H}_{s}(\mathbf{d})$ via its magnitude response $\mathbf{A}_{s}(\mathbf{d}) \in \mathbb{R}_+^{F \times 2}$ and ITD $\tau_{s}(\mathbf{d}) \in \mathbb{R}$.
The predicted HRTF magnitude can be converted to the time domain with the minimum phase~\cite{Kistler1992}, and ITD is compensated by shifting the time-domain impulse responses.

\subsection{Neural field (NF) for HRTF spatial upsampling}

An NF for HRTF spatial upsampling is trained to be a map from a sound source direction to a representation of the HRTF.
The HRTF field~\cite{Zhang2023} predicts the magnitude response $\mathbf{A}_s(\mathbf{d})$ as follows:
\begin{equation}
    \mathbf{A}_s(\mathbf{d}) = \texttt{NF}(\mathbf{d} \mid \bm{\Xi}_s),
    \label{eq:nfsingle}
\end{equation}
where \texttt{NF} is specialized for each subject $s$, and $\bm{\Xi}_s$ denotes its parameters.
Since HRTFs have similarities across different subjects, we can pre-train a generic NF for multiple subjects with a small number of subject-specific parameters~\cite{Zhang2023,niirf}:
\begin{equation}
    \mathbf{A}_s(\mathbf{d}) = \texttt{NF}(\mathbf{d} \mid \bm{\Gamma}, \bm{\Xi}_s),
    \label{eq:nfpeft}
\end{equation}
where $\bm{\Gamma}$ denotes the parameters shared across different subjects, and $\bm{\Xi}_s$ steers the generic model to represent the HRTF of a target subject $s$.
During its pre-training, \texttt{NF} is trained on given pairs of subject $s \in \mathcal{S}$ and direction $\mathbf{d} \in \mathcal{D}$, jointly optimizing $\bm{\Gamma}$ and $\bm{\Xi}_s, \forall s \in \mathcal{S}$.
Then, we adapt \texttt{NF} to a new target subject $\bar{s}$ by optimizing the subject-specific parameters $\bm{\Xi}_{\bar{s}}$ on the limited number of measurements, while freezing the shared parameters $\bm{\Gamma}$. 
This adaptation is more efficient than training a new NF in \eqref{eq:nfsingle} from scratch.
In \cite{niirf}, the low-rank adaptation (LoRA)~\cite{Hu2021} shows promising adaptation performance by modifying the weight of the $l$th layer $\mathbf{W}_l \in \mathbb{R}^{N_l \times M_l}$ as follows:
\begin{equation}
    \mathbf{W}_{l, s} = \mathbf{W}_l + \mathbf{u}_{l, s} \mathbf{v}_{l, s}^\mathsf{T},
    \label{eq:lora}
\end{equation}
where $(\cdot)^\mathsf{T}$ denotes the transpose, and $\mathbf{u}_{l, s} \in \mathbb{R}^{N_l}$ and $\mathbf{v}_{l, s} \in \mathbb{R}^{M_l}$ are subject-specific vectors for building a rank-$1$ matrix.

While existing NF-based methods have focused on approximating the HRTF magnitude, we predict ITD in sample $\tau_s(\mathbf{d})$ as well:
\begin{equation}
    \mathbf{A}_s(\mathbf{d}), \tau_s(\mathbf{d}) = \texttt{NF}(\mathbf{d} \mid \bm{\Gamma}, \mathbf{U}_s, \mathbf{V}_s),
    \label{eq:nfpeft-tau}
\end{equation}
where the subject-specific parameters $\bm{\Xi}_s$ in \eqref{eq:nfpeft} are split into $\mathbf{U}_s = [\mathbf{u}_{1, s}, \ldots, \mathbf{u}_{L, s}]$ and $\mathbf{V}_s= [\mathbf{v}_{1, s}, \ldots, \mathbf{v}_{L, s}]$ for LoRA.
This allows us to reconstruct the head-related impulse responses except for the distance-dependent time-of-arrival and to evaluate the impulse responses with the objective measures used in the listener acoustic personalization (LAP) challenge%
\footnote{\url{https://www.sonicom.eu/lap-challenge}}.

\section{Proposed method: RANF}
\label{sec:prop}

\subsection{Overview of RANF}
\label{sec:overview}

As \textcircled{\scriptsize 1} in Fig.~\ref{fig:overview}, we first retrieve $K$ subjects $r_k^s \in \mathcal{S}$ for a given target subject $s$, where $k = 1, \ldots, K$ is the index of the retrieved subjects.
The $K$ subjects are selected on the basis of the similarity of the ITD and/or HRTF magnitude to those of a target subject $s$ at the measured directions in $\mathcal{D}'$.
To avoid clutter, we omit the superscript $s$ in $r_k^s$ hereafter.
In \textcircled{\scriptsize 2}, we compute the HRTF magnitudes $\mathbf{A}_{r_{1:k}}(\mathbf{d})$ and ITDs $\tau_{r_{1:k}}(\mathbf{d})$ for the retrieved subjects at the desired direction $\mathbf{d}$.
These are fed into NF as additional clues to predict the HRTF of the target subject in \textcircled{\scriptsize 4}:
\begin{equation}
    \mathbf{A}_s(\mathbf{d}), \tau_s(\mathbf{d}) 
    = \texttt{RANF}(\mathbf{d}, \mathbf{A}_{r_{1:k}}(\mathbf{d}), \tau_{r_{1:k}}(\mathbf{d}) \mid
     \bm{\Gamma}, \mathbf{U}_s, \mathbf{V}_{r_{1:k}}),
    \label{eq:ranf}
\end{equation}
where subject-specific parameters are selected in \textcircled{\scriptsize 3}, combining parameters $\mathbf{U}_s$ for the target subject $s$ and parameters $\mathbf{V}_{r_{1:k}} = \{\mathbf{V}_{r_1}, \ldots, \mathbf{V}_{r_k}\}$ for the retrieved subjects $r_k$.
This allows RANF to perform retrieval-dependent processing.

During the pre-training of RANF, we optimize the shared parameters $\bm{\Gamma}$ and the subject-specific parameters $(\mathbf{U}_s, \mathbf{V}_{r_{1:K}})$ on given pairs of the target subject $s \in \mathcal{S}$ and sound source direction $\mathbf{d} \in \mathcal{D}$.
We retrieve the $K$ subjects also from $\mathcal{S}$, where we ensure $r_{k} \neq s$ for all $k$.
The shared and subject-specific parameters are optimized to minimize the sum of the log-spectral distortion (LSD) on the HRTF magnitudes and the $\epsilon$-insensitive mean absolute error (MAE) on the ITDs%
\footnote{
We compute LSD and the oracle ITD based on \texttt{Spatial Audio Metrics}:
\url{https://github.com/Katarina-Poole/Spatial-Audio-Metrics}.
}:
\begin{align}
    \mathcal{L}_{\text{pre-train}}
    &\!=\! \sum_{s \in \mathcal{S}} \sum_{\mathbf{d} \in \mathcal{D}} \mathcal{L}(\mathbf{H}_s^{\star}(\mathbf{d}), \mathbf{H}_s(\mathbf{d})), \\
    \!\!\mathcal{L}(\mathbf{H}^{\star}(\mathbf{d}), \mathbf{H}(\mathbf{d}))
    &\!=\! \texttt{LSD}(\mathbf{A}^\star(\mathbf{d}), \mathbf{A}(\mathbf{d})) \!+\! \lambda \texttt{MAE}_\epsilon(\tau^\star(\mathbf{d}), \tau(\mathbf{d})),\!\!
    \label{eq:loss}
\end{align}
where $\lambda \in \mathbb{R}_+$ is a hyperparameter, $(\cdot)^\star$ indicates the oracle value,
$\epsilon$ is set to $0.5$ to consider the quantization error in the oracle sample-level ITD,
and LSD is defined as
\begin{equation}
\texttt{LSD}(\mathbf{A}^\star(\mathbf{d}), \mathbf{A}(\mathbf{d}))
    = \frac{1}{2} \sum_{c=1}^2 \sqrt{ \frac{1}{F} \sum_{f=1}^F \Big(20 \log_{10} \frac{A_{c,f}(\mathbf{d})}{A_{c,f}^\star (\mathbf{d})}  \Big)^2}.
    \label{eq:mse}
\end{equation}
In \eqref{eq:loss} and \eqref{eq:mse}, we omit the subscript $s$ for the subject to show a general case,
and $c \in \{1, 2\}$ and $f = 1, \ldots, F$ are the channel and frequency indices, respectively.
Then, we adapt the pre-trained generic model to a new target subject $\bar{s} \notin \mathcal{S}$ by optimizing only $\mathbf{U}_{\bar{s}}$ to minimize the following loss function:
\begin{equation}
    \mathcal{L}_{\text{adaptation}}
    = \sum_{\mathbf{d} \in \mathcal{D}'} \mathcal{L}(\mathbf{H}_{\bar{s}}^{\star}(\mathbf{d}), \mathbf{H}_{\bar{s}}(\mathbf{d})).
\end{equation}
During the adaptation, we freeze the shared parameters $\bm{\Gamma}$ and the parameters relying on the retrieved subjects $\mathbf{V}_{r_{1:k}}$.

\begin{figure*}[t]
    \centering
    \includegraphics[width=0.96\linewidth]{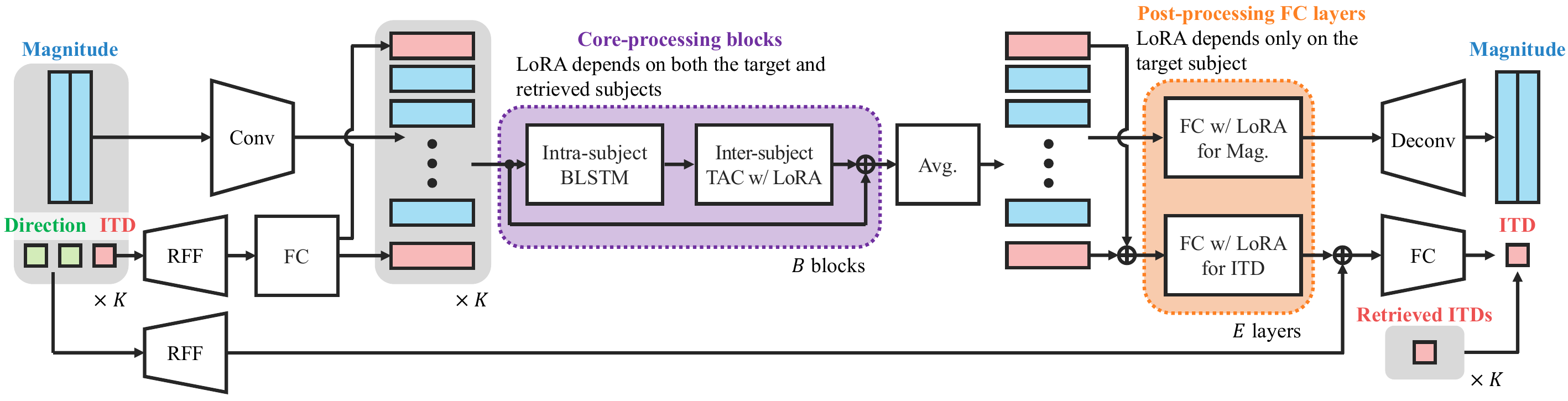}
    \vspace{-3mm}
     \caption{
       Network architecture for the proposed RANF.
       HRTF magnitudes are encoded by a convolution network and decoded by a deconvolution network.
       The retrieved ITDs are encoded as RFF with the sound source direction and passed to an FC layer.
       The feature sequence in the gray box is for each of the $K$ retrieved subjects and processed in parallel except for the inter-subject TAC module.
       The features are aggregated across all retrieved subjects by averaging (Avg.).
    }
    \label{fig:network}
    \vskip -3mm
\end{figure*}

\subsection{HRTF retrieval in \textcircled{\scriptsize 1} and \textcircled{\scriptsize 2}}
\label{sec:retrieval}

To efficiently personalize a generic NF to the target subject $s$, we retrieve $K$ subjects from a given HRTF dataset whose HRTF will be helpful to predict the HRTF of the target subject.
In detail, we retrieve top-$K$ subjects on the basis of a HRTF error measure $\mathcal{L}_\text{retrieval}$:
\begin{equation}
    \tilde{r}_1, \ldots, \tilde{r}_{K} \leftarrow \argmin_{r_{1:K} \in [\mathcal{S}]^{K}} \sum_{k=1}^{K} \sum_{\mathbf{d} \in \mathcal{D}'} \mathcal{L}_\text{retrieval}(\mathbf{H}_s(\mathbf{d}), \mathbf{H}_{r_k}(\mathbf{d})),
    \label{eq:retrieval-for-ranf}
\end{equation}
where $[\mathcal{S}]^{K}$ denotes the set of all subsets of $\mathcal{S}$ with $K$ distinct elements.
The choice of $\mathcal{L}_\text{retrieval}$ affects the prediction quality, and we investigate using LSD or MAE on ITDs in Section~\ref{sec:ablation}.

In \textcircled{\scriptsize 2}, the HRTF magnitudes and ITDs of the retrieved subjects $r_{1:K}$ at the desired direction $\mathbf{d}$ are computed, assuming that all of the desired directions are in the set of sound source directions $\mathcal{D}$.
This assumption is not overly limiting if the HRTF dataset for retrieval has sufficiently dense spatial grids.
If the desired direction $\mathbf{d}$ is not in $\mathcal{D}$, we can use another NF to predict the HRTFs at $\mathbf{d}$ for the subjects in the dataset. We shall leave this for future work.

\subsection{Network architecture for RANF in \textcircled{\scriptsize 4}}
\label{sec:network}

Figure~\ref{fig:network} illustrates a network architecture for RANF.
The HRTF magnitude of each retrieved subject is encoded with a 1D convolution network that downsamples the number of frequency bins $F$ to $F'$, where input and output channels are $2$ and $C$, respectively.
The convolution network captures the local frequency information in the retrieved HRTF magnitude.
Meanwhile, the retrieved ITD is transformed into random Fourier features (RFF)~\cite{Tancik2020} with the azimuth and elevation of the sound source.
The RFF is converted into two $C$-dimensional feature vectors by a fully connected (FC) layer and concatenated at both ends of the encoded magnitude.
Hence, ITD-related features are passed to the beginning of both forward and backward LSTMs in the following intra-subject BLSTM modules.
We obtain $K$ feature sequences with shape $(F'+2) \times C$.

Next, we feed the feature sequences into $B$ core-processing blocks, each consisting of an intra-subject BLSTM module and an inter-subject transform-average-concatenate (TAC) module~\cite{Luo2020icassp} with a residual connection.
In the intra-subject BLSTM module, we handle each of the $K$ feature sequences separately and perform sequence modeling along the $F'+2$ feature dimensions.
On the other hand, the inter-subject TAC module mixes the feature vectors across different retrieved subjects, where each of $F'+2$ feature dimensions is treated separately.
This module includes LoRA to perform retrieval and target-dependent processing.

After the $B$ core-processing blocks, we average the $K$ feature sequences into a single target feature sequence.
A portion of this sequence corresponding to the HRTF magnitude is processed by $E$ post-processing FC layers with LoRA, where both LoRA vectors are specified by the target subject as in \eqref{eq:lora}.
The output sequence is decoded to the two-channel HRTF magnitude by a 1D deconvolution block.
Meanwhile, the features at both ends are summed up for ITD prediction and passed to $E$ post-processing FC layers.
We then incorporate an additional RFF computed only from the desired direction $\mathbf{d}$ and predict the residual between the oracle ITD and the average ITD of the retrieved subjects via FC layers without LoRA.
That is, the network is trained to predict $\tau_s(\mathbf{d}) - (1/K) \sum_{k=1}^K \tau_{r_k} (\mathbf{d})$.

\subsection{Inter-subject TAC module}
\label{sec:tac}

For the inter-subject TAC module, we treat each of the $F'+2$ feature vectors in the sequence separately.
This module mixes the information across different retrieved subjects by using a TAC module~\cite{Luo2020icassp} and performs subject-specific processing with LoRA.

\begin{figure}[t]
    \centering
    \includegraphics[width=0.99\linewidth]{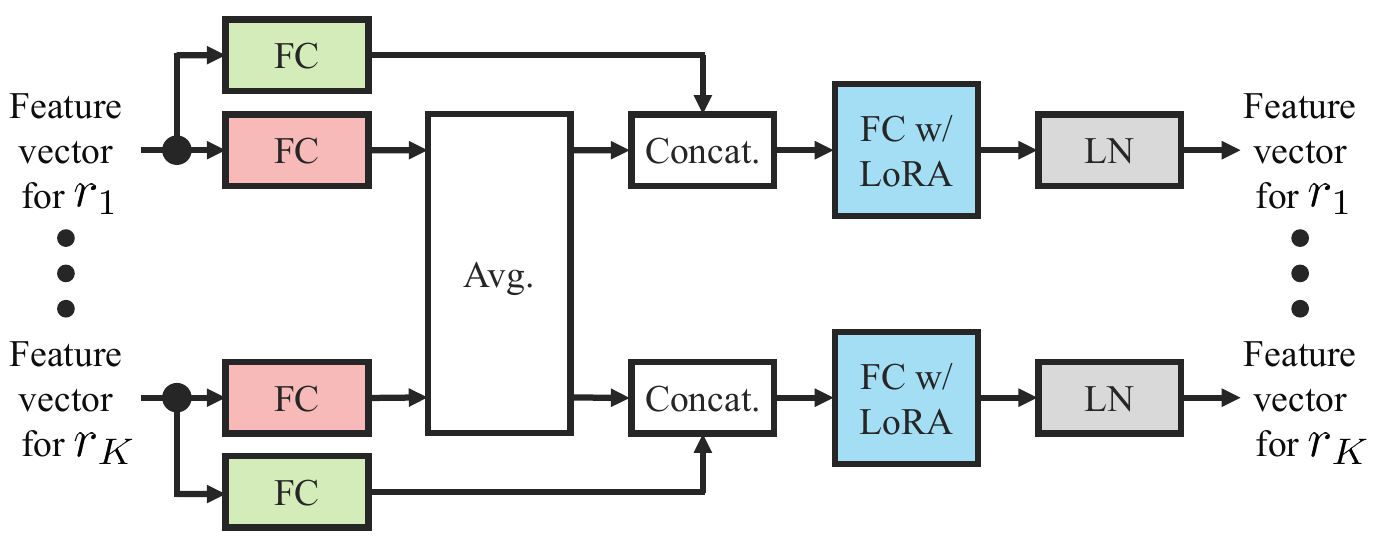}
    \vskip -3mm
    \caption{
        Inter-subject TAC module.
        Trainable layers are colored, where the layers with the same color share their parameters.
    }
    \label{fig:loratac}
    \vskip -3mm
\end{figure}

Figure~\ref{fig:loratac} shows the inter-subject TAC module, where LN is layer normalization.
The input feature vectors are processed by two FC layers, where the outputs of one FC layer are averaged over the retrieved subjects.
The averaged vector is concatenated with the outputs of the other FC layer.
This TAC operation was originally proposed for multi-channel speech separation to handle multi-channel features in a permutation invariant manner~\cite{Luo2020icassp}.
We use the TAC operation to efficiently mix the features for multiple retrieved subjects.
It can scale the model to an arbitrary number of retrieved subjects, which is advantageous for environments with variable computational resources.

The concatenated features are then passed to an FC layer with LoRA.
Here, LoRA differs from \eqref{eq:lora} as follows:
\begin{equation}
    \mathbf{W}_{b, s, k} = \mathbf{W}_b + \mathbf{u}_{b, s} \mathbf{v}_{b, r_k}^\mathsf{T},
    \label{eq:raglora}
\end{equation}
where $b = 1, \ldots, B$ indexes the core-processing blocks.
With the modification in \eqref{eq:raglora}, we aim to take the relation between the target and retrieved subjects into account.
Since we retrieve $r_k$ from the set of pre-training subjects $\mathcal{S}$ even in inference, $\mathbf{v}_{b, r_k}$ can be pre-trained and fixed during the adaptation.

\section{Experiments}
\label{sec:exp}

\begin{table*}[t]
\sisetup{
detect-weight,
mode=text,
tight-spacing=true,
round-mode=places,
round-precision=1,
table-format=1.1,
table-number-alignment=center
}
  \caption{LSD, ILD error, and ITD error for different numbers of measured directions for adaptation.}
  \label{tab:main-results}
  \vskip -0.1in
  \centering
  \resizebox{\textwidth}{!}{
  \begin{tabular}{l*{3}{S[table-format=3.1]SS}S[table-format=2.1]SS}
    \toprule
     & \multicolumn{3}{c}{3 measurements} & \multicolumn{3}{c}{5 measurements}
     & \multicolumn{3}{c}{19 measurements} & \multicolumn{3}{c}{100 measurements} \\
    \cmidrule(lr){2-4} \cmidrule(lr){5-7} \cmidrule(lr){8-10} \cmidrule(lr){11-13}
    Methods & {ITD [$\mu$s]} & {ILD [dB]} & {LSD [dB]} & {ITD [$\mu$s]} & {ILD [dB]} & {LSD [dB]} & {ITD [$\mu$s]} & {ILD [dB]} & {LSD [dB]} & {ITD [$\mu$s]} & {ILD [dB]} & {LSD [dB]} \\
    \midrule
    Nearest neighbor & 274.2 & 7.6 & 8.7 & 154.2 & 4.8 & 8.3 & 108.3 & 3.0 & 5.4 & 44.0 & 1.4 & 3.4 \\
    HRTF selection (ITD) & 26.3 & 1.4 & 6.4 & 24.5 & 1.5 & 6.4 & 23.2 & 1.6 & 6.7 & 20.0 & 1.4 & 6.3 \\
    HRTF selection (LSD) & 37.5 & 1.5 & 5.8 & 35.7 & 1.4 & 5.6 & 37.4 & 1.5 & 5.5 & 31.4 & 1.3 & 5.4 \\
    \midrule
    NF (CbC)~\cite{Zhang2023} & 22.1 & 1.5 & 4.9 & 20.7 & 2.0 & 5.2 & 14.8 & 1.8 & 5.0 & 11.8 & 1.7 & 5.1 \\
    NF (LoRA)~\cite{niirf} & 28.6 & 1.3 & 4.7 & 24.7 & 1.4 & \bfseries 4.6 & 14.7 & 1.1 & 4.1 & \bfseries 9.1 & 1.1 & 3.8 \\
    \midrule
    RANF (proposed) & \bf20.5 & \bfseries 1.2 & \bfseries 4.4 & \bfseries 18.7 & \bfseries 1.2 & \bfseries 4.6 & \bfseries 14.2 & \bfseries 1.0 & \bfseries 3.6 & 10.0 & \bfseries 0.8 & \bfseries 3.0 \\
    \bottomrule
  \end{tabular}
  }
  \vskip -4mm
\end{table*}

\subsection{Dataset and experimental setup}

We evaluate RANF on the SONICOM HRTF dataset~\cite{engel2023}, which includes 200 subjects and 793 measurements per subject.
HRTFs were sampled at 48 kHz, and a free-field compensation was performed with a minimum-phase filter as in the LAP challenge.
The first 179 subjects were used for pre-training, and the remaining 20 subjects were used to evaluate the adaptation capability.
We removed subject \texttt{P0079} from the dataset due to its atypical ITD behavior.
Within the 179 subjects, the last 19 subjects were used for validation.
The upsampling performance was investigated on four different sets $\mathcal{D}'$ of measured directions for the target subjects, where the number of measured directions varied in $(3, 5, 19, 100)$ as in the LAP challenge.

The convolution and deconvolution networks consisted of 4 layers with the PReLU activation, where the encoded dimension $C$ was 128.
The number of core-processing blocks $B$ was $4$, and the BLSTM had $64$ units for each direction in each block.
The FC layers in the inter-subject TAC module used the GELU activation.
The number of post-processing FC layers $E$ was set to $2$, and the last ITD prediction network consisted of $2$ layers.
In \eqref{eq:loss}, we set $\lambda$ to $20.8$ to change the loss scale from sample to microsecond.
The neural network was first pre-trained for 200 epochs and then adapted for 500 epochs by using the RAdam optimizer with an initial learning rate of $0.001$.
The learning rate was multiplied by $0.9$ when the validation loss did not decrease for 10 successive epochs.
The model with the best validation loss was used for adaptation.
During adaptation, we did not perform validation to exploit all the measurements for adaptation.
Our training and inference scripts are available online%
\footnote{\url{https://github.com/merlresearch/ranf-hrtf}}.

We adopted two baselines: the nearest neighbor direction and HRTF selection.
The best-fitting subject was selected from the 179 subjects in terms of LSD or MAE of ITDs.
In addition, we compared the proposed method to NFs with two kinds of subject-specific parameters: the conditioning by concatenation (CbC)~\cite{Zhang2023} and LoRA~\cite{niirf}.
CbC introduces a subject-specific latent vector and concatenates it to the input.
Its dimension was set to $32$ following the original paper.

\begin{figure}[t!]
\centering
\includegraphics[width=0.95\columnwidth]{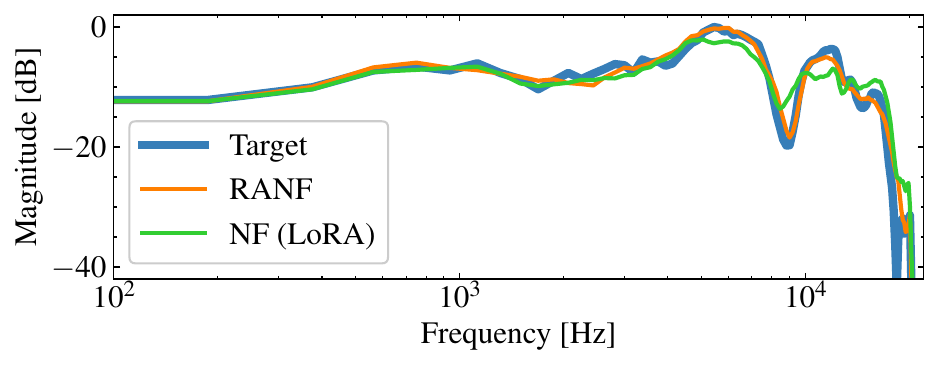}
\vskip -3mm
\caption{Ipsilateral HRTF magnitudes for a subject with  $(\theta, \phi) = (\pi/2, 0)$.}
\label{fig:hrtf-magnitude}
 \vskip -3mm
\end{figure}

\subsection{Comparison with existing methods}

Table~\ref{tab:main-results} shows the average scores used in the LAP challenge, where we evaluate the performance only on the unmeasured directions $\mathbf{d} \in \mathcal{D} \backslash \mathcal{D}'$.
In RANF, we used MAE on ITDs to retrieve $5$ subjects based on a preliminary study.
The nearest neighbor direction performed poorly with only 3 or 5 measurements, which highlights the difficulty of HRTF spatial upsampling from such a limited number of measurements.
Meanwhile, HRTF selection retained reasonable performance even in the sparsest condition.

Regarding the existing NF-based methods, both methods substantially improved LSD from HRTF selection.
NF (LoRA), which contains more subject-specific parameters, achieved better LSD but increased the ITD error for 3 and 5 measurements, likely because LoRA overfits for such a limited number of measurements.
By incorporating the top-$5$ retrievals, RANF consistently outperformed the baselines and improved upsampling performance from NF (LoRA) except for the ITD with 100 measurements.
Figure~\ref{fig:hrtf-magnitude} shows examples of the predicted HRTF magnitude for subject \texttt{P0181} under the 100 measurements condition.
We can observe that RANF captured the peak and notch around 10 kHz more accurately than NF (LoRA).

\begin{table}[t]
\sisetup{
detect-weight,
mode=text,
tight-spacing=true,
round-mode=places,
round-precision=1,
table-format=2.1,
table-number-alignment=center
}
  \caption{Upsampling results with different retrieval strategies.}
  \label{tab:retrievals}
  \vskip -0.1in
  \centering
  \resizebox{\linewidth}{!}{
  \begin{tabular}{cS[round-precision=0,table-format=2]S*{2}{S[table-format=1.1]}}
    \toprule
      Retrieval criterion & {\# of retrievals $ K$} & {ITD [$\mu$s]} & {ILD [dB]} & {LSD [dB]} \\
     \midrule
     \multirow{3}{*}{Random} & 1 & 26.6 & 1.3 & \bfseries 4.4 \\
     & 5 & 23.8 & 1.3 & 4.5 \\
     & 10 & 23.5 & 1.3 & \bfseries 4.4 \\
     \midrule
     \multirow{3}{*}{LSD} & 1 & 29.9 & 1.3 & 4.6 \\
     & 5 & 22.4 & \bfseries 1.2 & \bfseries 4.4 \\
     & 10 & 22.7 & \bfseries 1.2 & \bfseries 4.4 \\
     \midrule
     \multirow{3}{*}{ITD error} & 1 & 23.8 & \bfseries 1.2 & \bfseries 4.4 \\
     & 5 & \bfseries 20.5 & \bfseries 1.2 & \bfseries 4.4 \\
     & 10 & 22.7 & 1.3 & \bfseries 4.4 \\
    \bottomrule
  \end{tabular}
  }
  \vskip -3mm
\end{table}

\begin{table}[t]
\sisetup{
detect-weight,
mode=text,
tight-spacing=true,
round-mode=places,
round-precision=1,
table-format=2.1,
table-number-alignment=center
}
  \caption{Results with and without post-processing FC layers.
  Params refers to the number of shared trainable parameters.
  }
  \label{tab:lora-balance}
  \vskip -0.1in
  \centering
  \resizebox{\linewidth}{!}{
  \begin{tabular}{cccS*{2}{S[table-format=1.1]}}
    \toprule
     \begin{tabular}{c}Post-processing \\ FC layers \end{tabular} & $B$ & Params (M) & {ITD [$\mu$s]} & {ILD [dB]} & {LSD [dB]} \\
     \midrule
      \ding{53} & 4 & 0.75 & 24.6 & 1.4 & 4.7 \\
      \ding{53} & 5 & 0.89 & 24.3 & 1.3 & 4.7 \\
     \midrule
     \ding{51} & 4 & 0.82 & \bfseries 20.5 & \bfseries 1.2 & \bfseries 4.4 \\
    \bottomrule
  \end{tabular}
  }
  \vskip -3mm
\end{table}

\subsection{Ablation study}
\label{sec:ablation}

To show the efficacy of our retrieval strategy and network architecture, we perform an ablation study on the 3 measurement condition.

\noindent \textbf{Retrieval strategy}:
In RANF, we need to choose the HRTF error measure $\mathcal{L}_\text{retrieval}$ in \eqref{eq:retrieval-for-ranf}.
In addition to top-$K$ retrieval in terms of LSD and the ITD error, we evaluated the random retrieval of subjects for pre-training.
Table~\ref{tab:retrievals} presents the HRTF spatial upsampling performance with different retrieval criteria and different numbers of retrievals.
The results show comparable LSD and ILD error regardless of the number of retrievals or their criterion.
Conversely, ITD upsampling performance was substantially improved when performing retrieval based on the ITD error.
It also helped to retrieve more than one subject, and we ultimately found that retrieving 5 subjects with the ITD-based criterion leads to the best results.

\noindent \textbf{Efficacy of post-processing FC layers}:
Our NF shown in Fig.~\ref{fig:network} leverages LoRA in two formulations.
LoRA in the core-processing blocks depends on both target and retrieved subjects as in \eqref{eq:raglora}, while LoRA in the post-processing FC layers relies solely on the target subject as in \eqref{eq:lora}.
Table~\ref{tab:lora-balance} presents the upsampling performance with and without the post-processing FC layers, where $K=5$.
It can be observed that the post-processing FC layers bring better performance even with fewer shared trainable parameters.

\section{Conclusion}
\label{sec:conclusion}

We present RANF, an HRTF spatial upsampling method that enhances NF by leveraging HRTF selection.
RANF retrieves multiple subjects from a given HRTF dataset, and our experiments demonstrated the benefit of multiple retrievals in terms of ITD error.
The winning system of the LAP challenge 2024 is based on RANF, and our future work will include a detailed analysis of RANF after the organizer releases the baselines and full datasets.

\clearpage
\balance
\bibliographystyle{IEEEtran}
\bibliography{refs}

\end{document}